\documentclass[final,authoryear,3p,times,twocolumn]{elsarticle}

\usepackage{amssymb}

\usepackage[nodots]{numcompress}

\usepackage{epsfig}

\journal{Advances in Space Research}

\begin{document}

\begin{frontmatter}

\title{Testing the No-Hair Theorem with Observations of Black Holes in the Electromagnetic Spectrum}

\author{Tim Johannsen$^1$ and Dimitrios Psaltis$^2$}

\address{$^1$Physics Department, University of Arizona, 1118 E. 4th Street, Tucson, AZ 85721, USA; timj@physics.arizona.edu \\
$^2$Astronomy Department, University of Arizona, 933 N.\ Cherry Ave., Tucson, AZ 85721, USA; dpsaltis@email.arizona.edu}

\begin{abstract}

According to the no-hair theorem, astrophysical black holes are uniquely described by their mass and spin. In this paper, we review a new framework for testing the no-hair hypothesis with observations in the electromagnetic spectrum. The approach is formulated in terms of a Kerr-like spacetime containing a quadrupole moment that is independent of both mass and spin. If the no-hair theorem is correct, then any deviation from the Kerr metric quadrupole has to be zero. We show how upcoming VLBI imaging observations of Sgr A* as well as spectroscopic observations of iron lines from accreting black holes with IXO may lead to the first astrophysical test of the no-hair theorem.

\end{abstract}


\end{frontmatter}

\section{Introduction}

The no-hair theorem encapsulates the unique property of black holes in general relativity that these objects are completely and uniquely described by their masses and spins (Israel 1967, 1968; Carter 1971, 1973; Hawking 1972; Robinson 1975). This theorem relies on the cosmic censorship conjecture stating that naked singularities have to be enclosed by an event horizon (Penrose 1969; see, however, Shapiro et al. 1995) and on the causality requirement that the exterior spacetime is free of closed timelike curves. Under these assumptions, all astrophysical black holes should be described by the Kerr metric.

While there has been, to date, a wealth of observations arguing in favor of the existence of black holes (e.g., Tremaine et al. 2002; Sch\"odel et al. 2002; McClintock \& Remillard 2006; Ghez et al. 2008; Gillessen et al. 2009), there is no direct evidence yet for an actual event horizon, and various alternative explanations have been suggested (e.g., Friedberg, Lee, \& Pang 1987; Manko \& Novikov 1992; Mazur \& Mottola 2001; Barcel\'{o} et al. 2008; Yunes \& Pretorius 2009; see also Psaltis et al. 2008).

It is therefore incumbent to test the no-hair theorem observationally. The detection of gravitational waves from extreme mass-ratio inspirals with the {\it Laser Interferometer Space Antenna} (LISA) will be able to probe the immediate vicinity of supermassive black holes (Ryan 1995, 1997a, 1997b; Barack \& Cutler 2004, 2007; Collins \& Hughes 2004; Glampedakis \& Babak 2006; Gair et al.\ 2008; Li \& Lovelace 2008; Apostolatos et al.\ 2009; Vigeland \& Hughes 2010). Alternative tests include observations of stars on orbits very close to the galactic center (Will 2008; Merritt et al. 2010) as well as of pulsar black-hole binaries (Wex \& Kopeikin 1999).

In Johannsen \& Psaltis (2010a, 2010b), we analyzed a framework for performing a test of the no-hair theorem with observations of black holes in the electromagnetic spectrum. We used a quasi-Kerr metric (Glampedakis \& Babak 2006) that parametrizes a potential deviation from the Kerr metric at the quadrupole order. We identified imaging observations of black holes using very long baseline interferometry (VLBI), as well as observations of relativistically broadened iron lines with the {\it International X-Ray Observatory} (IXO) and of continuum disk spectra as promising avenues for such tests.

Sgr A*, the center of the Milky Way, is the ideal candidate for VLBI imaging. This supermassive black hole has the largest angular size of any black hole in the sky (e.g., Psaltis 2008). In addition, its emission at sub-millimeter wavelengths becomes optically thin (e.g., Broderick et al. 2009) and is no longer obscured by interstellar scattering (e.g., Bower et al. 2006). Finally, such observations are technically feasible (Doeleman et al. 2008; Fish \& Doeleman 2009).

\section{Properties of Quasi-Kerr Black Holes}

\begin{figure*}[t]
\begin{center}
\psfig{figure=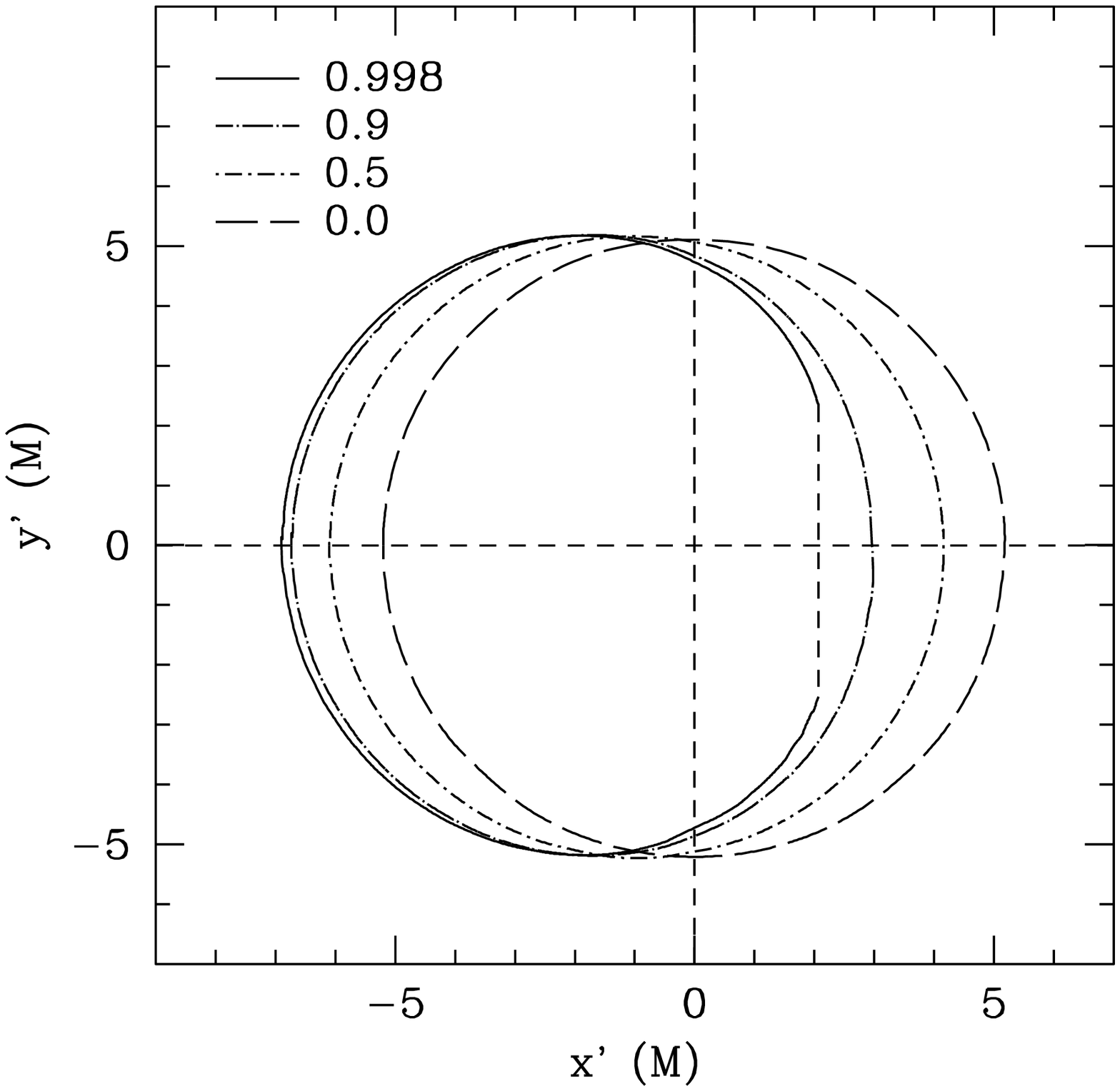,height=3in}
\psfig{figure=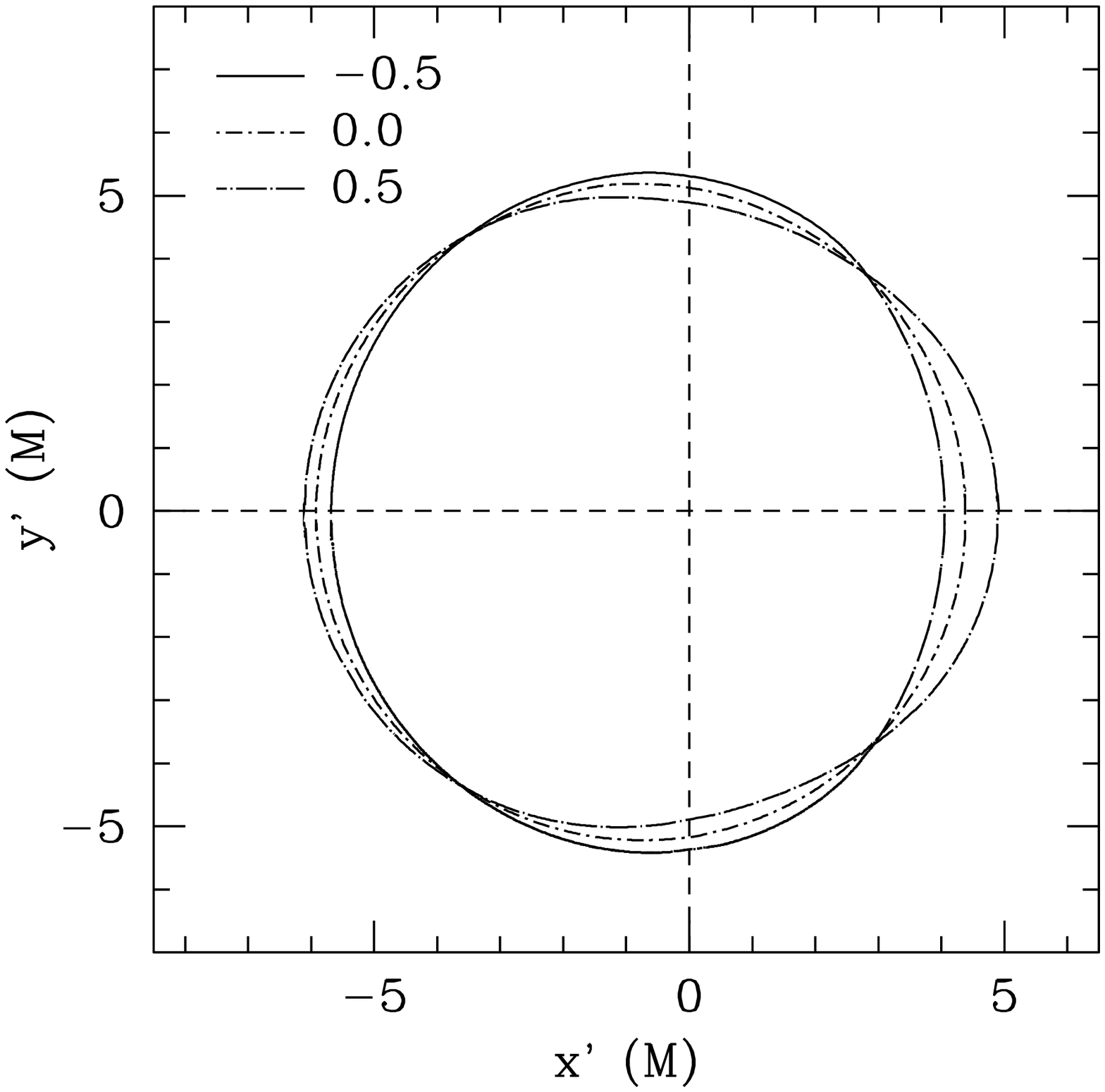,height=3in}
\end{center}
\caption{Images of rings of light of (left) a Kerr and (right) a quasi Kerr black hole at an inclination $\cos i=0.25$. Increasing values of the spin lead to a displacement of the ring in the image plane, but the ring remains (nearly) circular for values of the spin $a\lesssim0.9M$ (left panel). Nonzero values of the quadrupolar parameter $\epsilon$ (right panel) cause the ring image to become asymmetric (Johannsen \& Psaltis 2010b).}
\label{}
\end{figure*}

The quasi-Kerr metric of Glampedakis \& Babak (2006) parametrizes a deviation from the Kerr metric by an independent quadrupole moment. In this parametrization, the quadrupole moment $Q$ takes the form (Glampedakis \& Babak 2006)
\begin{equation}
Q = -M(a^2+\epsilon M^2),
\end{equation}
where $M$ is the mass of the black hole, $a=J/M$ is its spin per unit mass, and $\epsilon$ is a dimensionless parameter expressing the deviation from the Kerr quadrupole moment. For a Kerr black hole, $\epsilon=0$. This approach is valid for slowly spinning black holes with values of the spin $a\lesssim0.4M$ (Glampedakis \& Babak 2006; Johannsen \& Psaltis 2010a).

According to the no-hair theorem, mass and spin are the only free parameters of a black hole, i.e., $\epsilon=0$. Therefore, with a measurement of the mass, spin, and quadrupole moment of a black hole we can test the no-hair theorem. This framework allows either for a null-hypothesis test of the no-hair theorem within general relativity (Collins \& Hughes 2004; Hughes 2006) or for a test of general relativity itself in the strong-field regime (Johannsen \& Psaltis 2010a). If a deviation of a multipole moment from the Kerr metric is measured, i.e., in our case a nonzero value of the quadrupolar parameter $\epsilon$, there are two possibilities. If general relativity is assumed to be correct, then the compact object has to be of a different kind other than a black hole. On the other hand, if the object is otherwise known to possess an event horizon, then both the no-hair theorem and general relativity are incorrect. However, a non-detection of a deviation from the Kerr metric alone cannot give a definite answer regarding the validity of general relativity and the no-hair theorem, because a large class of alternative theories of gravity likewise predicts Kerr-type black holes (Psaltis et al. 2008).

As in Johannsen \& Psaltis (2010a), we term a black hole that is described by the quasi-Kerr metric a quasi-Kerr black hole, similarly to the term bumpy black hole (Collins \& Hughes 2004; Vigeland \& Hughes 2010). A deviation of one of the multipole moments from their respective value in the Kerr metric changes the spacetime properties of quasi-Kerr black holes. These, in turn, directly affect observables. In Johannsen \& Psaltis (2010a), we identified a number of spacetime properties that are critical for a measurement of the black hole parameters mass, spin, and quadrupole moment, and we analyzed the dependence of these properties of the parameter $\epsilon$.

The most important properties of quasi-Kerr black holes are the following: (i) The location of the innermost stable circular orbit (ISCO). This orbit is often taken to be the inner edge of the accretion disk of a black hole. For increasing values of the parameter $\epsilon$, the ISCO is pushed to larger radii. (ii) The redshift of photons emitted from the accretion disk. For a photon emitted by a particle on the ISCO and observed at infinity, increasing the value of the spin augments the observed photon redshift, while increasing the value of the parameter $\epsilon$ decreases the observed redshift. (iii) The radius of the circular photon orbit. The projected location of this orbit along null geodesics marks the outer edge of the shadow of a black hole (Bardeen 1973). Similarly to the location of ISCO, increasing the value of the parameter $\epsilon$ shifts the circular photon orbit to larger radii. (iv) Gravitational lensing. Close to the black hole the lensing of photons is either increased or decreased, depending on the value of the parameter $\epsilon$ and the direction of the spin (Johannsen \& Psaltis 2010a).

\section{Testing the No-Hair Theorem Observationally}

By measuring the mass, spin, and quadrupole moment of a black hole, we are in a position to test the no-hair theorem observationally. This can be done in terms of several observables which have the potential to be realized in the near future.

\subsection{Rings of Light}

\begin{figure*}[t]
\begin{center}
\psfig{figure=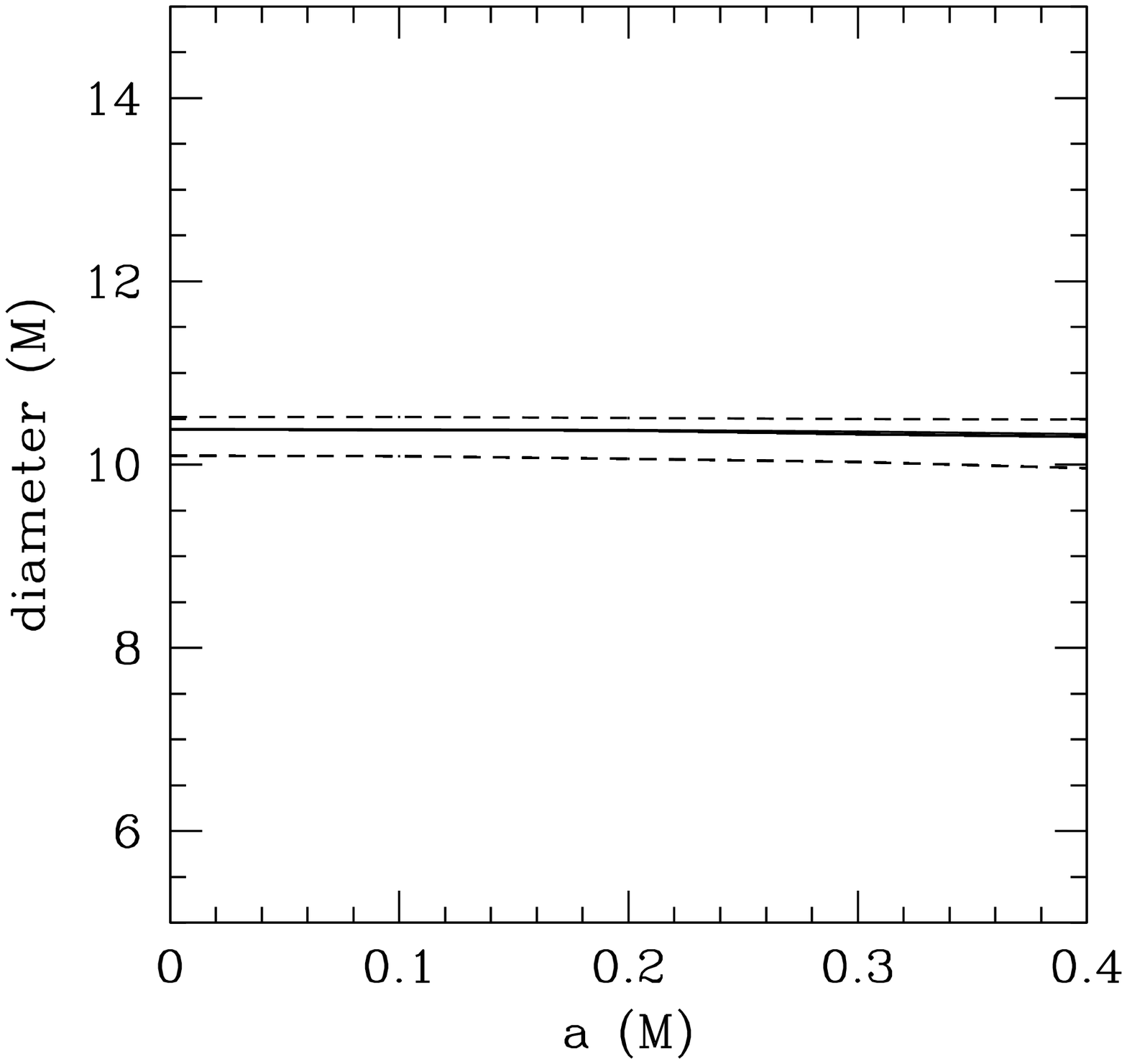,height=2.1in}
\psfig{figure=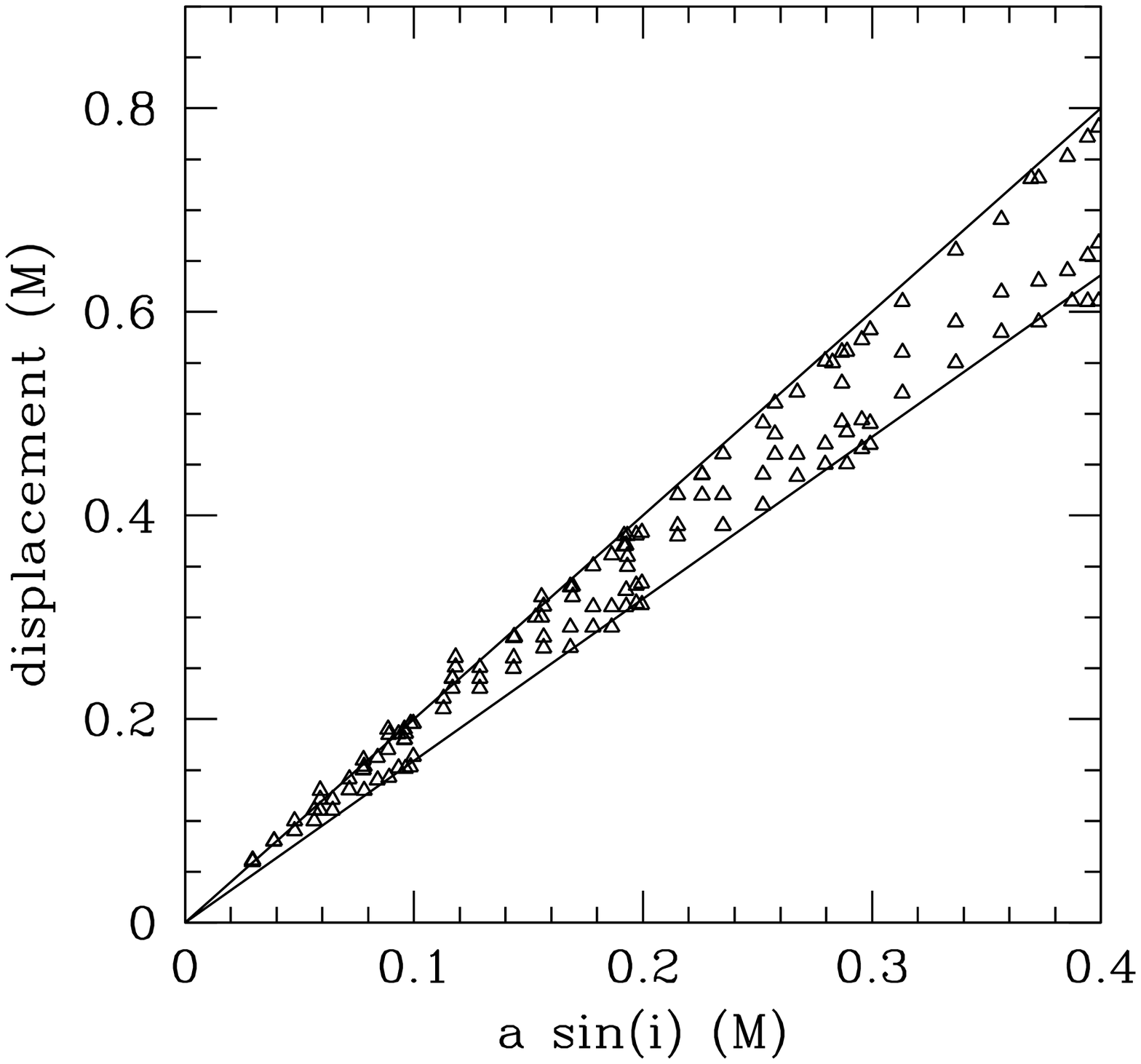,height=2.1in}
\psfig{figure=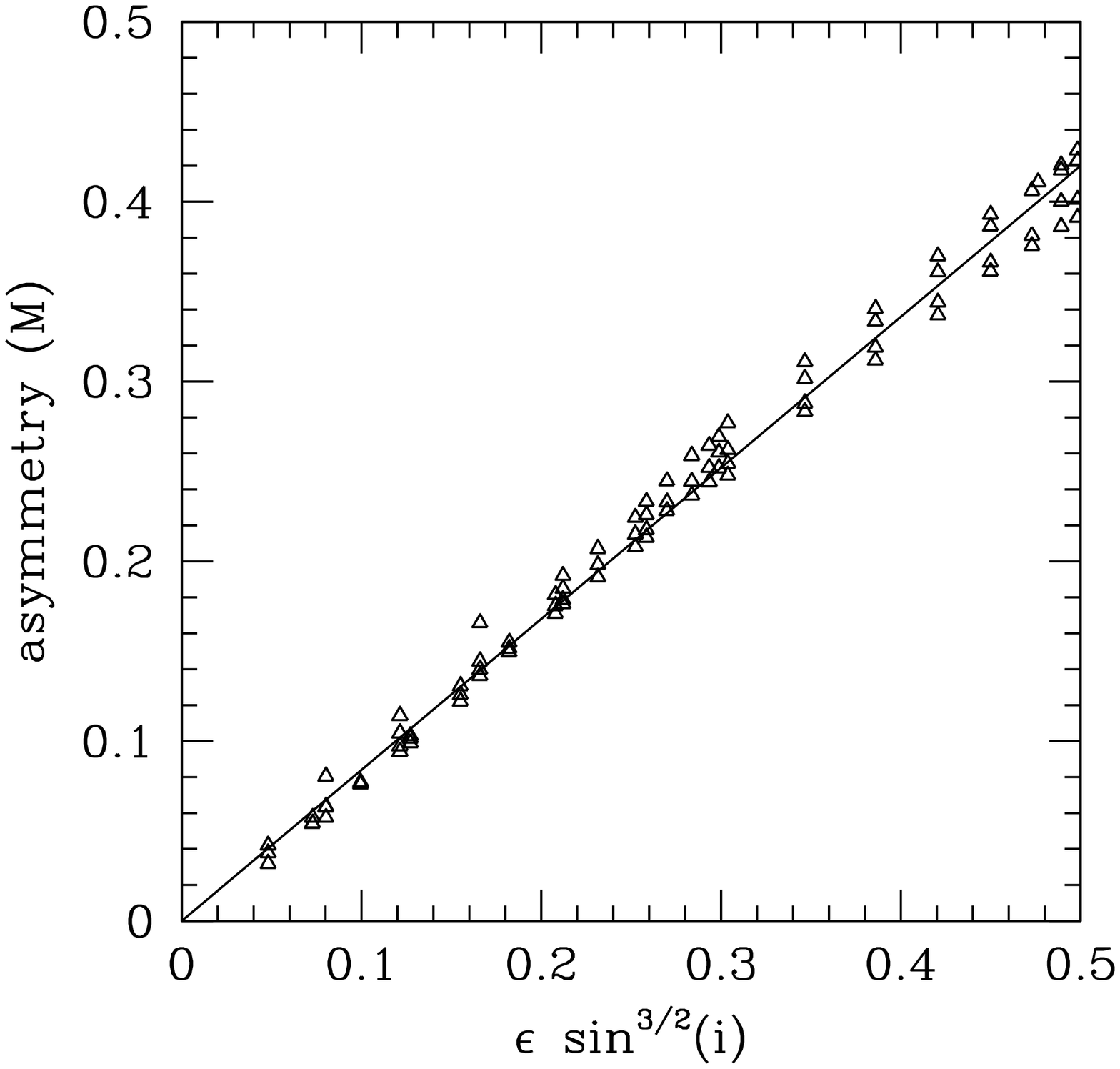,height=2.1in}
\end{center}
\caption{Left: The ring diameter versus spin for inclinations $17^\circ\leq i \leq86^\circ$ for a Kerr black hole (solid lines) and for a quasi-Kerr black hole with a value of the parameter $\epsilon=0.5$ (dashed lines). The diameter is practically independent of the spin, inclination, and quadrupole moment with a constant value of $\simeq$10.4$M$ for a Kerr black hole. Center: The displacement of the ring of light as a function of $a\sin i$ for various values of the parameter $0\leq\epsilon\leq0.5$. The displacement depends only weakly on the parameter $\epsilon$. Right: The ring asymmetry versus $\epsilon\sin^{3/2} i$ for various inclinations $17^\circ\leq i \leq86^\circ$ and $0.0\leq a/M \leq0.4$. The asymmetry is nearly independent of the spin and hence provides a direct measure of a violation of the no-hair theorem (Johannsen \& Psaltis 2010b).}
\label{}
\end{figure*}

Imaging observations of a black hole are expected to reveal its shadow which can be used to infer the mass, spin, and inclination of the black hole (e.g., Falcke et al. 2000a; Broderick \& Loeb 2005, 2006; Fish \& Doeleman 2009). In addition to these parameters, the shape of the shadow also depends uniquely on the value of the quadrupole moment (Johannsen \& Psaltis 2010b). In practice, however, these measurements will be model dependent (e.g., Broderick \& Loeb 2009) and affected by finite telescope resolution (e.g., Falcke et al. 2000a; Takahashi 2004). Thus, additional observations might be required such as a multiwavelength study of polarization (Broderick \& Loeb 2006; see also Schnittman \& Krolik 2009, 2010).

In an optically thin accretion flow, photons can orbit around the black hole several times before they are detected at infinity producing a ring that can be significantly brighter than the underlying flow due to their long optical path (Bardeen 1973; Cunningham 1976; Luminet 1979; Laor, Netzer, \& Piran 1990; Viergutz 1993; Bao, Hadrava, \& {\O}stgaard 1994; ${\rm \check{C}ade\check{z}}$, Fanton, \& Calvani 1998; Falcke et al. 2000b; Agol \& Krolik 2000; Beckwith \& Done 2005; Johannsen \& Psaltis 2010b). The shape and location of this ``ring of light'' depends directly on the mass, spin, and quadrupole moment of the black hole. Therefore, the promise for extracting these parameters from a measurement of the ring is high (see Figure~1).

In Johannsen \& Psaltis 2010b, we explored in detail the dependence of images of rings of light on the mass, spin, inclination, and quadrupole moment of the black hole. The diameter of the ring of light as observed by a distant observer depends predominantly on the mass of the black hole and is nearly constant for all values of the spin and disk inclination as well as for small deviations of the quadrupole moment from the Kerr value. The spin of the black hole causes the ring to be displaced off center in the image plane with only a weak dependence on the parameter $\epsilon$. In all cases, the ring of a Kerr black hole remains nearly circular except for very large values of the spin $a\gtrsim0.9M$. However, if $\epsilon\neq0$, the ring becomes asymmetric in the image plane. This asymmetry is a direct measure for a violation of the no-hair theorem (see Figure~2).

\subsection{Relativistically Broadened Iron Lines}

\begin{figure}[t]
\begin{center}
\includegraphics[width=0.45\textwidth]{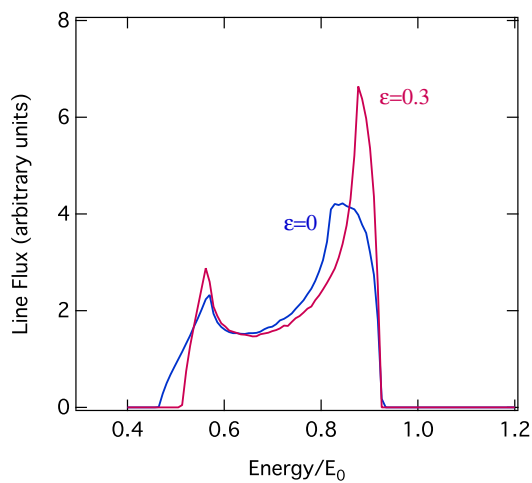}
\end{center}
\caption{Profiles of relativistically broadened iron lines for two values of the quadrupole deviation parameter $\epsilon$. The other black hole parameters are spin $a=0.4M$ and inclination $i=45^\circ$ with an emissivity profile $\propto~1/r^2$.}
\label{}
\end{figure}

Another observable are relativistically broadened iron lines emitted from the accretion disk of a black hole. The shape of these lines depends on the parameters of the black hole as well as on the geometry and emissivity profile of the accretion flow. The low-energy edge of such a line is determined by the location of the ISCO allowing for a measurement of the black hole spin (e.g., Reynolds \& Nowak 2003; Fabian 2007; Nandra et al. 2006; Miller 2007; Brenneman \& Reynolds 2009). In our framework, the line shape as well as the position of the ISCO likewise depend on the value of the quadrupole moment. Increasing the value of the parameter $\epsilon$ while fixing all other parameters leads to a narrower line shape (see Figure~3).

\section{Conclusions}

Observations of astrophysical black holes open up a new window for testing one of the most intriguing predictions of general relativity, the no-hair theorem. In this paper, we reviewed a framework that allows us to perform such tests, and we identified two key observables.

Sub-millimeter VLBI is expected to be able to image the accretion flow around Sgr A* resolving the shadow of the black hole (Doeleman et al. 2008; Fish \& Doeleman 2009). The  ring of light of Sgr A* with its relatively large size and especially its high brightness depends directly on the mass, spin, quadrupole moment, and inclination of the central black hole, and its properties are (nearly) independent of the details of the accretion flow geometry. Therefore, the brightness profile of the ring remains constant over an otherwise turbulent accretion flow which makes it the ideal target for VLBI observations that by design require long integration times.

Relativistically broadened iron lines will be observed with IXO providing an additional approach to testing the no-hair theorem. These observations will be complementary to the mapping of black hole spacetimes with LISA (see Hughes 2006) and the tracing of stellar orbits in close proximity to the galactic center (Will 2008; Merritt et al. 2010).


\begin{thebibliography}{00}

\bibitem[Agol \& Krolik (2000)]{agol00} Agol, E. \& Krolik, J. H. 2000, ApJ 528, 161-170
\bibitem[Apostolatos et al. (2009)]{apo09} Apostolatos, T.~A., Lukes-Gerakopoulos, G. \&, Contopoulos, G. 2009, Phys. Rev. Lett., 103, 111101(4)
\bibitem[Bao et al. (1994)]{bao94} Bao, G., Hadrava, P., \& {\O}stgaard, E. 1994, ApJ, 435, 55-65
\bibitem[Barack \& Cutler (2004)]{bar04} Barack, L., \& Cutler, C. 2004, Phys. Rev. D, 69, 082005(24)
\bibitem[Barack \& Cutler (2007)]{bar04} Barack, L., \& Cutler, C. 2007, Phys. Rev. D, 75, 042003(11)
\bibitem[Barcel\'o et al. (2008)]{barc08} Barcel\'o, C., Liberati, S., Sonego, S., \& Visser, M. 2008, Phys. Rev. D, 77, 044032(13)
\bibitem[Bardeen (1973)]{bard73} Bardeen, J. M. 1973, in Black Holes, ed. C. DeWitt \& B. S. DeWitt (New York: Gordon and Breach)
\bibitem[Beckwith \& Done (2005)]{beck05} Beckwith, K. \& Done, C. 2005, MNRAS 359, 1217-1228
\bibitem[Bower et al. (2006)]{bow06} Bower, G. C., Goss, W. M., Falcke, H., Backer, D. C., \& Lithwick, Y. 2006, ApJ, 648, L127-L130
\bibitem[Brenneman \& Reynolds (2009)]{bre09} Brenneman, L. W. \& Reynolds, C. S. 2009, ApJ, 702, 1367-1386
\bibitem[Broderick \& Loeb(2005)]{bro05} Broderick, A. E. \& Loeb, A. 2005, MNRAS, 363, 353-362
\bibitem[Broderick \& Loeb(2006)]{bro06} Broderick, A. E. \& Loeb, A. 2006, MNRAS, 367, 905-916
\bibitem[Broderick \& Loeb (2009)]{brodlo09} Broderick, A. E. \& Loeb, A. 2009, ApJ 697, 1164-1179
\bibitem[Broderick et al. (2009)]{brodfi09} Broderick, A. E., Fish, V. L., Doeleman, S. S., \& Loeb, A. 2009, ApJ, 697, 45-54
\bibitem[${\rm \check{C}ade\check{z}}$ et al. (1998)]{cha98} ${\rm \check{C}ade\check{z}}$, A., Fanton, C., \& Calvani, M. 1998, New Astronomy, 3, 647-654
\bibitem[Carter (1971)]{cart71} Carter, B. 1971, Phys. Rev. Lett. 26, 331-333
\bibitem[Carter (1973)]{cart73} Carter, B. 1973, in Black Holes, ed. C. DeWitt \& B. S. DeWitt (New York: Gordon and Breach)
\bibitem[Collins \& Hughes (2004)]{coll04} Collins, N. A. \& Hughes, S. A. 2004, Phys. Rev. D 69, 124022(16)
\bibitem[Cunningham (1976)]{cun76} Cunningham, C. 1976, ApJ, 208, 534-549
\bibitem[Doeleman et al. (2008)]{doele08} Doeleman, S. S. et al. 2008, Nature, 455, 78-80
\bibitem[Fabian (2007)]{fab07} Fabian, A. C. 2007, in IAU Symp. 238, Black Holes: From Stars to Galaxies, ed. V. Karas \& G. Matt (Cambridge: Cambridge Univ. Press)
\bibitem[Falcke et al. (2000a)]{falk00a} Falcke, H., Melia, F., \& Agol, E. 2000a, ApJ 528, L13-L16
\bibitem[Falcke et al. (2000b)]{falk00b} Falcke, H., Melia, F., \& Agol, E. 2000b, in Cosmic Explosions: Tenth Astrophysical Conference, ed. S. S. Holt \& W. W. Zhang (Melville: American Inst. of Physics)
\bibitem[Fish \& Doeleman (2009)]{fish09} Fish, V. L. \& Doeleman, S. S. 2009, in IAU Symp. 261, Relativity in Fundamental Astronomy: Dynamics, Reference Frames, and Data Analysis, ed. S. Klioner, P. K. Seidelmann, \& M. Soffel (Cambridge: Cambridge Univ. Press)
\bibitem[Friedberg et al. (1987)]{fried87} Friedberg, R., Lee, T. D., \& Pang, Y. 1987, Phys. Rev. D 35, 3640-3657
\bibitem[Gair et al. (2008)]{gair08} Gair, J. R., Li, C., \& Mandel, I. 2008, Phys. Rev. D 77, 024035(23)
\bibitem[Ghez et al. (2008)]{ghez08} Ghez, A. M. et al. 2008, ApJ, 689, 1044-1062
\bibitem[Gillessen et al. (2009)]{gill09} Gillessen, S., Eisenhauer, F., Trippe, S., Alexander, T., Genzel, R., Martins, F., \& Ott, T. 2009, ApJ, 692, 1075-1109
\bibitem[Glampedakis \& Babak (2006)]{glam06} Glampedakis, K. \& Babak, S. 2006, Class. Quantum Grav. 23, 4167-4188
\bibitem[Hawking (1972)]{haw72} Hawking, S. W. 1972, Commun. Math. Phys. 25, 152-166
\bibitem[Hughes (2006)]{hug06} Hughes, S. A. 2006, AIP Conf. Proc. 873, 233-240
\bibitem[Israel (1967)]{isr67} Israel, W. 1967, Phys. Rev. 164, 1776-1779
\bibitem[Isreal (1968)]{isr68} Israel, W. 1968, Commun. Math. Phys. 8, 245-260
\bibitem[Johannsen \& Psaltis (2010a)]{joh10a} Johannsen, T., \& Psaltis, D. 2010a, ApJ, 716, 187-197
\bibitem[Johannsen \& Psaltis (2010b)]{joh10b} Johannsen, T., \& Psaltis, D. 2010b, ApJ, 718, 446-454
\bibitem[Laor et al. (1990)]{laor90} Laor, A., Netzer, H., \& Piran, T. 1990, MNRAS, 242, 560-569
\bibitem[Li \& Lovelace (2008)]{li08} Li, C., \& Lovelace, G. 2008, Phys. Rev. D, 77, 064022(10)
\bibitem[Luminet (1979)]{lum79} Luminet, J.-P. 1979, A\&A, 75, 228-235
\bibitem[Manko \& Novikov (1992)]{man92} Manko, V. S. \& Novikov, I. D. 1992, Class. Quantum Grav. 9, 2477-2487
\bibitem[Mazur \& Mottola (2001)]{maz01} Mazur, P. O. \& Mottola, E. 2001, arXiv:0109035
\bibitem[McClintock \& Remillard (2006)]{mcc06} McClintock, J. E. \& Remillard, R. A. 2006, in Compact Stellar X-Ray Sources, ed. W. H. G. Lewin \& M. van der Klis (Cambridge: Cambridge Univ. Press)
\bibitem[Merritt et al. (2010)]{mer10} Merritt, D., Alexander, T., Mikkola, S., \& Will, C. M. 2010, Phys. Rev. D, 81, 062002(17)
\bibitem[Miller (2007)]{mil07} Miller, J. M. 2007, ARA\&A, 45, 441-479
\bibitem[Nandra et al. (2006)]{nan06} Nandra, K., O'Neill, P. M., George, I. M., Reeves, J. N., \& Turner, T. J. 2006, Astron. Nachr., 327, 1039-1042
\bibitem[Penrose (1969)]{pen69} Penrose, R. 1969, Riv. del Nouvo Cimento, 1, 252-276
\bibitem[Psaltis (2008)]{psa08} Psaltis, D. 2008, Living Rev. Rel., 11, 9(61)
\bibitem[Psaltis et al. (2008)]{psa08} Psaltis, D., Perrodin, D., Dienes, K. R., \& Mocioiu, I. 2008, Phys. Rev. Lett. 100, 091101(4)
\bibitem[Reynolds \& Nowak (2003)]{rey03} Reynolds, C. S. \& Nowak , M. A. 2003, Phys. Rep., 377, 389-466
\bibitem[Robinson (1975)]{rob75} Robinson, D. C. 1975, Phys. Rev. Lett. 34, 905-906
\bibitem[Ryan (1995)]{rya95} Ryan, F. D. 1995, Phys. Rev. D 52, 5707(12)
\bibitem[Ryan (1997a)]{rya97a} Ryan, F. D. 1997a, Phys. Rev. D 56, 1845(11)
\bibitem[Ryan (1997b)]{rya97b} Ryan, F. D. 1997b, Phys. Rev. D 56, 7732(8)
\bibitem[Schnittman \& Krolik (2009)]{schn09} Schnittman, J. D. \& Krolik, J. H. 2009, ApJ, 701, 1175-1187
\bibitem[Schnittman \& Krolik (2010)]{schn10} Schnittman, J. D. \& Krolik, J. H. 2010, ApJ, 712, 908-924
\bibitem[Sch\"odel et al. (2002)]{scho02} Sch\"odel, R. et al. 2002, Nature, 419, 694-696
\bibitem[Shapiro et al. (1995)]{sha95} Shapiro, S. L., Teukolsky, S. A., \& Winicour, J. 1995, Phys. Rev. D, 52, 6982-6987
\bibitem[Takahashi (2004)]{taka04} Takahashi, R. 2004, ApJ 611, 996-1004
\bibitem[Tremaine et al. (2002)]{tre02} Tremaine, S., et al. 2002, ApJ, 574, 740-753
\bibitem[Viergutz (1993)]{vier93} Viergutz, S. U. 1993, Astron. Astrophys., 272, 355-377
\bibitem[Vigeland \& Hughes (2010)]{vig10} Vigeland, S. J. \& Hughes, S. A., 2010, Phys. Rev. D, 81, 024030(20)
\bibitem[Wex \& Kopeikin (1999)]{wex99} Wex, N., \& Kopeikin, S. M. 1999, ApJ, 514, 388-401
\bibitem[Will (2008)]{will08} Will, C. M.\ 2008, ApJ, 674, L25-L28
\bibitem[Yunes \& Pretorius (2009)]{yun09} Yunes, N. \& Pretorius, F. 2009, Phys. Rev. D 79, 084043(14)


\end{thebibliography}
\end{document}